# Gate depletion of an InSb two-dimensional electron gas

M. M. Uddin,[1] H. W. Liu,[1,2,3,a)] K. F. Yang,[1,2] K. Nagase,[1,2] K. Sekine,[1] C. K. Gaspe,[4] T. D. Mishima,[4] M. B. Santos,[4] and Y. Hirayama[1,2,5,b)]

[1]*Department of Physics, Tohoku University, Sendai, Miyagi 980-8578, Japan*

[2]*ERATO Nuclear Spin Electronics Project, Sendai, Miyagi 980-8578, Japan*

[3]*State Key Laboratory of Superhard Materials and Institute of Atomic and Molecular Physics, Jilin University, Changchun 130012, People's Republic of China*

[4]*Homer L. Dodge Department of Physics and Astronomy, University of Oklahoma, 440 West Brooks, Norman, Oklahoma 73019-2061, USA*

[5]*WPI Research Center, Advanced Institute for Materials Research, Tohoku University, Sendai 980-8577, Japan*

We investigated the gate control of a two-dimensional electron gas (2DEG) confined to InSb quantum wells with an $Al_2O_3$ gate dielectric formed by atomic layer deposition on a surface layer of $Al_{0.1}In_{0.9}Sb$ or InSb. The wider bandgap of $Al_{0.1}In_{0.9}Sb$ compared to InSb resulted in a linear, sharp, and non-hysteretic response of the 2DEG density to gate bias in the structure with an $Al_{0.1}In_{0.9}Sb$ surface layer. In contrast, a nonlinear, slow, and hysteretic (nonvolatile-memory-like) response was observed in the structure with an InSb surface layer. The 2DEG with the $Al_{0.1}In_{0.9}Sb$ surface layer was completely depleted by application of a small gate voltage (~ -0.9 V).

A split-gate technique is now widely used for the fabrication of semiconductor nanostructures (e.g., quantum point contacts and quantum dots): a negative bias on surface gates locally depletes the underlying two-dimensional electron gas (2DEG) confined to a quantum well (QW), leaving electrons only in a desired region surrounded by gate areas. This technique has already been applied to most semiconductors (Si, GaAs, InAs, InGaAs, etc.), but not to the typical narrow-gap semiconductor InSb even though InSb is particularly appealing for spintronics applications (spin field-effect transistors,[1-3] heat-driven spin devices,[4] electron or nuclear-spin-based quantum bits,[5,6] etc.) and for the detection of signatures of majorana Fermions due to strong spin-orbit coupling and giant $g$-factor.[7] Because the relatively low barrier height of a Schottky contact to (Al)InSb produces high current leakage,[8,9] top gating with gate dielectrics[10] becomes an attractive alternative for surface-gate fabrication of InSb QWs. However, several studies have highlighted difficulties in growing high quality gate dielectrics on (Al)InSb and in forming a good interface between the dielectric and (Al)InSb layers.[3,8]

More recently, we have grown a high quality $Al_2O_3$ gate dielectric on an InSb QW structure with an InSb surface layer (hereafter referred to as "sample 1") using atomic layer deposition (ALD).[11] The Fermi level of this sample is tuned almost entirely across the band gap of InSb via gate bias. The good interface between the $Al_2O_3$ and InSb layers makes this possible, and allows us to study the importance of the layer sequence in gate controllability. However, the 2DEG in such a QW cannot be fully depleted because hole accumulation at the InSb surface layer screens the gate electric field. Based on a self-consistent Schrödinger-Poisson (SP) simulation, we predicted that the 2DEG would be depleted in an InSb QW with a wider-band-gap $Al_xIn_{1-x}Sb$ surface layer that is expected to prevent the hole accumulation.[11] In this letter, we show experimental evidence that the 2DEG in an InSb QW with an $Al_{0.1}In_{0.9}Sb$ surface layer (sample 2) is completely depleted. Particularly noteworthy is that the $Al_{0.1}In_{0.9}Sb$ layer also has the advantages of suppressing a parallel conduction channel and of keeping a relatively low interface trap density in the depletion process as revealed by our modified SP simulation. The success in gate depletion of the 2DEG enables the application of the split-gate technique to InSb QWs, which has significant implications for realizing InSb-based spintronics.

The layer structure of sample 2, shown on the bottom axis of Fig. 1(a), is the same as that of sample 1 except for the absence of an InSb cap layer and for a shorter distance between the second silicon (Si) δ-doped layer and the QW (15 nm for sample 1 and 10 nm for sample 2). After a Hall bar mesa (80 μm ×30 μm) was defined by photolithography, a 40-nm-thick $Al_2O_3$ layer was deposited using ALD at 130°C without any surface treatment[12] (details of the ALD growth and the epitaxial layer structure are described in Ref. 11). The lower panel of Fig. 1 shows atomic force microscopy (AFM) images of ALD-$Al_2O_3$ grown on the InSb (sample 1) and $Al_{0.1}In_{0.9}Sb$ (sample 2) surface layers with a root-mean-square (rms) roughness of 0.57 nm and 0.33 nm, respectively, suggesting a smooth and dense insulator film for both samples. Indium was then evaporated for the Ohmic contacts and the top gate. We note that the following results were also obtained for a sample with $Al_2O_3$ grown by ALD at 150°C. All experiments and simulations in this work were carried out at a temperature of 2 K unless noted otherwise. A standard AC lock-in technique (13.3 Hz; 35 nA) was used for both Hall and magnetoresistance (MR) measurements.

Figure 1(a) shows the energy band diagram (up to 85 nm in depth) of sample 2 at zero gate bias ($V_g$) calculated from the SP simulation with a Schottky barrier model.[11,13] In this SP simulation, a unique fitting parameter is the Schottky barrier height $\phi_B$ that is determined by properties of the semiconductor surface and interface states. The energy difference between the conduction band (CB) minimum $E_c$ at zero depth and the Fermi energy $E_F$ (set at 0 eV) corresponds to $\phi_B$, which can be adjusted manually for agreement between the simulated and the measured 2DEG density $n_s$. For instance, $\phi_B = 0.14$ eV in Fig. 1(a) gives $n_s = 3.32 \times 10^{15}$ m$^{-2}$, consistent with the Hall measurement result at $V_g = 0$. It is also seen that such a $\phi_B$ locates the CB discontinuity at the $Al_{0.1}In_{0.9}Sb/Al_{0.2}In_{0.8}Sb$ interface a little above $E_F$, thus preventing the formation of a parallel conduction channel at the discontinuity.[11] This is evidenced by the zero-bias MR plot in the upper inset to Fig. 1(b): the MR property is similar to that of a single 2DEG except for the presence of non-zero longitudinal resistance $R_{xx}$ under the Hall ($R_{xy}$) plateau, indicating that the parallel channel is strongly suppressed. Note that the parallel channel emerges at a small $V_g \sim 0.05$ V (data not shown) because the positive bias will lower the CB discontinuity below $E_F$. More importantly, the relatively wide band gap ($\sim 0.42$ eV)[14] of $Al_{0.1}In_{0.9}Sb$ in sample 2 keeps the valence band (VB) maximum $E_v$ at zero depth far from $E_F$ as shown in Fig. 1(a), which avoids hole accumulation at the surface under negative bias[11] and thereby enables complete depletion of the 2DEG. Fig. 1(b) shows the $V_g$ dependence of $n_s$ and the mobility ($\mu$) of sample 2, obtained from Hall and resistivity measurements. Apparently, both $n_s$ and $\mu$ decrease with increasing $|V_g|$. The data at $V_g = -0.6$ V gives $R_{xx} \sim 23$ kΩ at zero magnetic field (see also the MR curve in Fig. 1(b), lower inset). A further increase of $|V_g|$ greatly enhances $R_{xx}$ and thus invalidates the Hall measurement, as expected when both $n_s$ and $\mu$ become sufficiently small. The intersection of the $n_s$-$V_g$ plot with the horizontal axis at $n_s = 0$ in Fig. 1(b) allows us to estimate a pinch-off voltage of $V_p \sim -0.9$ V for complete depletion of the 2DEG. We should mention that $|V_p|$ is much smaller than a threshold voltage of 4 V for current leakage through the ALD-grown $Al_2O_3$ in this sample. Thus, gate depletion of the InSb 2DEG has been demonstrated in sample 2.

Besides its role in gate depletion of the InSb 2DEG discussed above, the $Al_{0.1}In_{0.9}Sb$ surface layer is also found to form a good interface with the ALD-grown $Al_2O_3$ layer. For comparison, the $n_s$-$V_g$ plots of sample 1 and sample 2 are shown in Fig. 2(a). It is clear that the change of $n_s$ in response to $V_g$ is different for the two samples. Firstly, $n_s$ of sample 1 decreases slowly and nonlinearly with increasing $|V_g|$ and finally tends to saturate, while that of sample 2 decreases rapidly and almost linearly with increasing $|V_g|$, resulting in a large slope of $dn_s/dV_g = 3.9 \times 10^{15}$ m$^{-2}$V$^{-1}$. Secondly, a notable hysteresis is observed in sample 1 but only a very small one appears in sample 2. We applied both an equivalent capacitance model and a SP simulation to explain these differences. The inset to Fig. 2(a) depicts an equivalent capacitance circuit for sample 2, where $C_{ox}$ and $C_{sc}$ are the capacitance per unit area of the $Al_2O_3$ and semiconductor layers above the QW, respectively, and $C_{it} = e^2 D_{it}$ ($e$ is the electron charge and $D_{it}$ is the interface trap density) is the capacitance associated with the interface traps[15] and $C_{2D} = e^2 m^* m_e / \pi \hbar^2$ ($m^* \sim 0.0135$ is the effective mass of InSb, $m_e$ is the electron mass, and $\hbar$ is Planck's constant divided by $2\pi$) is the quantum capacitance[16] of the InSb 2DEG. In the ideal case of $C_{it} = 0$, the $n_s$-$V_g$ dependence is given by $en_s = C_{tot}V_g$ ($C_{tot}$ is the total capacitance of $C_{ox}$, $C_{sc}$ and $C_{2D}$), as indicated by a dashed line in Fig. 2(a). The slope $dn_s/dV_g \sim 5.5 \times 10^{15}$ m$^{-2}$V$^{-1}$ of this line is close to that of sample 2, indicating a good interface between the ALD-grown $Al_2O_3$ and the $Al_{0.1}In_{0.9}Sb$ layer. In the case of $C_{it} \neq 0$, $dn_s/dV_g = [1 + C_{it}(C_{sc} + C_{2D})/(C_{sc}C_{2D})]^{-1} C'_{tot}/e$ ($C'_{tot}$ is the total capacitance of $C_{ox}$, $C_{sc}$, $C_{it}$ and $C_{2D}$) is deduced from the equivalent circuit and used to fit our data, giving a relatively low $D_{it} = 1.05 \times 10^{16}$ m$^{-2}$eV$^{-1}$. However, this capacitance model is only appropriate for the data analysis at $-1.6$V $\leq V_g \leq -1.2$V in sample 1 because it is difficult to account for the effects of the parallel conduction channel at $V_g > -1.2$V and hole accumulation at $V_g < -1.6$V (see below) in the equivalent circuit. The value of $dn_s/dV_g = 6 \times 10^{14}$ m$^{-2}$V$^{-1}$ at $-1.6$V $\leq V_g \leq -1.2$V (downward sweep) yields a high $D_{it} = 1.7 \times 10^{17}$ m$^{-2}$eV$^{-1}$. Although $D_{it}$ of the two samples is quite different, care should be taken in concluding that the interface of $Al_2O_3/Al_{0.1}In_{0.9}Sb$ is better than that of $Al_2O_3$/InSb because $D_{it}$ is energy dependent, as discussed below.

Fig. 2(b) shows the $V_g$ dependence of $E_F$ obtained from the SP simulation. Since $E_F$ is fixed at 0 eV in the simulation, the modification of $E_F$ is reflected in the movement of $E_v$ relative to $E_F$. The energy difference $E_F$-$E_v$ is calculated by $E_g$-$E_c$, where $E_g$ is the band gap of the semiconductor surface layer and $E_c$ is determined in the same way as $\phi_B$. It is shown in Fig. 2(b) that $E_F$ in sample 2 is readily tuned by a small bias, indicating a very weak pinning of $E_F$. Because our SP simulation is invalid in the presence of a parallel channel or hole accumulation, there are only two data points for sample 1. The data point at $V_g$ = -1.6V suggests that $E_v$ touches the Fermi level. As $|V_g|$ is further increased, holes accumulate at the semiconductor surface and screen the external electric field, which accounts for the saturation of $n_s$ in Fig. 2(a). On the other hand, $E_F$-$E_v$ of sample 2 at $V_p$ = -0.9V (solid symbol) is greater than zero, indicating that $n_s$ continues to decrease with increasing $|V_g|$ until depletion. In order to analyze the interface trap states, we extend the SP simulation to include a model of metal-oxide-semiconductor (MOS) structures. In this modified SP simulation, the $V_g$ applied to the metal causes voltage drops across the oxide layer and at the semiconductor surface. The voltage $|V_{ox}|$ across the oxide is equal to $|V_g|$-$E_c$/e, neglecting any work-function difference and interfacial layer thickness. The Gauss law then gives $Q_{it}$ =$\varepsilon_0\varepsilon_{ox}E_{ox}$-$\varepsilon_0\varepsilon_{sc}E_{sc}$ (electric field $E_{ox}$= $V_{ox}/d_{ox}$ with $Al_2O_3$ thickness $d_{ox}$ = 40 nm, electric field $E_{sc}$ at the semiconductor surface obtained from the SP simulation). Based on the equation $dQ_{it}/d(E_F$-$E_v)$ =-e$D_{it}$,[15] we obtain the energy dependent $D_{it}$ of the two samples shown in the inset to Fig. 2(b). We see that $D_{it}$ around the midgap (0.21 eV) of $Al_{0.1}In_{0.9}Sb$ in sample 2 is about $1\times10^{16}$ m$^{-2}$eV$^{-1}$, consistent with the result of the capacitance model. It is clear that a low $D_{it}$ is kept in the depletion process, accounting for a low-voltage gate operation. Because $D_{it}$ around the midgap (0.118 eV) of InSb in sample 1 cannot be calculated due to the presence of a parallel channel, we cannot compare the interfacial properties of the two samples. Nevertheless, the data point indicating $D_{it}$ ~$1.6\times10^{17}$ m$^{-2}$eV$^{-1}$ near the valence band of sample 1 agrees with the capacitance model calculation.

We now discuss the density hysteresis observed in sample 1. This hysteresis is found to behave like a nonvolatile memory, as shown in Fig. 3: it depends on the sweep direction and range of $V_g$ but is nearly independent of the sweep rate, which is different from the observations in InAs/AlGaSb[17] or HgTe/HgCdTe[18] QWs. Such a hysteretic behaviour is reminiscent of the capacitance-voltage (C-V) characteristic of nonvolatile $Al_2O_3$ memory devices, where charge traps in a nonstoichiometric $Al_2O_3$ layer dominate the memory effect.[19] Here we also assign charge traps in the $Al_2O_3$ layer to account for our hysteresis. As is shown in Fig. 3, the hysteresis becomes more prominent at more negative bias, where $D_{it}$ is large and holes accumulate at the semiconductor surface layer as discussed above. We expect that the oxygen vacancies in the $Al_2O_3$ layer with charge-state switching levels[20] near the bandgap of InSb might act as charge traps: the fully ionized oxygen vacancies trap (detrap) electrons (holes) in the downward sweep of $V_g$ and the neutral ones detrap (trap) electrons (holes) in the upward sweep of $V_g$ via a tunnelling process. The stored charge $Q_{tr}$ in the oxide, proportional to $D_{it}$, the hole density $Q_h$ and the oxygen vacancy density, induces a voltage drop that determines the hysteresis width. A large change in $V_g$ leads to a large change in $Q_h$ and $D_{it}$, accordingly resulting in a large hysteresis. Note that such a hysteresis can be observed up to 100 K.

In conclusion, we have demonstrated gate depletion of the 2DEG in InSb QWs with an $Al_{0.1}In_{0.9}Sb$ surface layer. This surface layer forms an excellent interface with an $Al_2O_3$ gate dielectric layer grown by ALD at low temperature (130°C) in the absence of any surface treatment, leading to excellent gate performance characterized by low-voltage operation and negligible hysteresis. Following this development, the design and fabrication of split-gate-defined InSb quantum point contacts are now in progress.

M.M.U. would like to thank IIARE, WPI-AIMR, Tohoku University for S-SDC and RA fellowship. H.W.L. thanks the Program for New Century Excellent Talents of the University in China.

a)Electronic email: liuhw@m.tohoku.ac.jp

b)Electronic email: hirayama@m.tohoku.ac.jp

**FIG. 1.** (Color online) (a) Band profile of sample 2 vs. depth along the growth direction at zero bias calculated from a self-consistent Schrödinger-Poisson simulation. The Fermi energy $E_F$ is fixed at 0 eV, and $E_c$ and $E_v$ denote the conduction-band minimum and the valence-band maximum, respectively. The Si δ-doped regions in the $Al_xIn_{1-x}Sb$ layers (see bottom axis) are indicated by vertical arrows. (b) Electron density $n_s$ and mobility $\mu$ of the InSb 2DEG as a function of gate bias $V_g$ for sample 2. The dash-dotted line points to the pinch-off voltage $V_p$. Insets show the magnetic-field ($B$) dependent longitudinal resistance $R_{xx}$ and Hall resistance $R_{xy}$ at $V_g = 0$ V and -0.6 V. Lower panel: AFM images of the ALD-grown $Al_2O_3$ of sample 1 and sample 2.

**FIG. 2.** (Color online) (a) $n_s$-$V_g$ plots of sample 1 and sample 2. Arrows indicate the gate sweep direction. Inset shows an equivalent capacitance circuit of the gated InSb QW. The symbol $C_x$ denotes the capacitance per unit area. The calculation of $C_{ox(sc)} = \varepsilon_{ox(sc)}\varepsilon_0/d_{ox(sc)}$ ($\varepsilon_0$ is vacuum permittivity) is based on the following parameters: dielectric constant $\varepsilon$ ($Al_2O_3$:7; InSb:16.82; $Al_{0.1}In_{0.9}Sb$:16.34; $Al_{0.2}In_{0.8}Sb$:15.86) and thickness $d$ in units of nm ($Al_2O_3$:40; InSb:10; $Al_{0.1}In_{0.9}Sb$:30; $Al_{0.2}In_{0.8}Sb$:20). The $n_s$-$V_g$ dependence of the capacitance circuit with $C_{it} = 0$ is indicated by the dashed line. (b) $V_g$ dependence of $E_F$-$E_v$ calculated from the SP simulation. The open symbols are calculations for the data in Fig. 2(a) (downward sweep) and the solid one is for the data at $V_p$ = -0.9V in Fig. 1(b). The dashed line is drawn to guide the eye. Inset shows the interface trap density $D_{it}$ as a function of $E_F - \bar{E}_V$ obtained from our modified SP simulation. $\bar{E}_V$ is the average value of $E_F$-$E_v$ at neighbouring $V_g$.

**FIG. 3.** (Color online) Hysteresis in the $n_s$-$V_g$ plot of sample 1 as a function of gate sweep direction and range. The solid (dashed) line denotes the upward (downward) sweep.

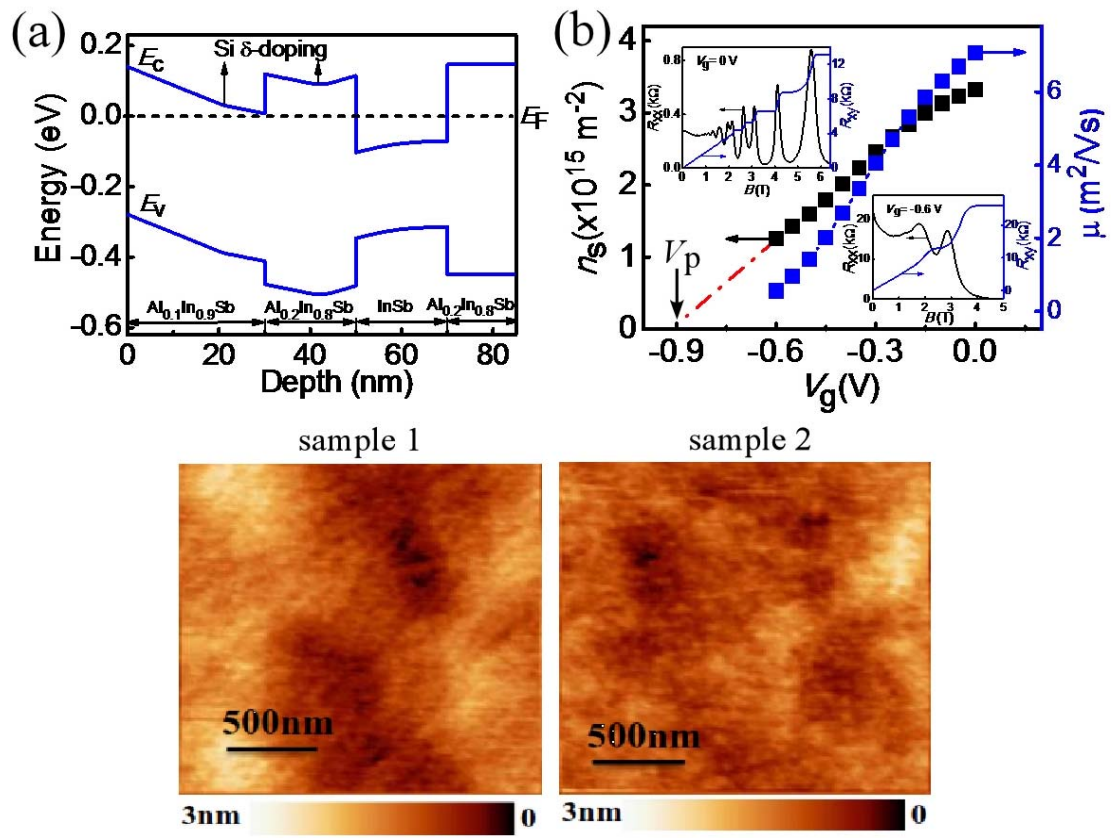

FIG. 1

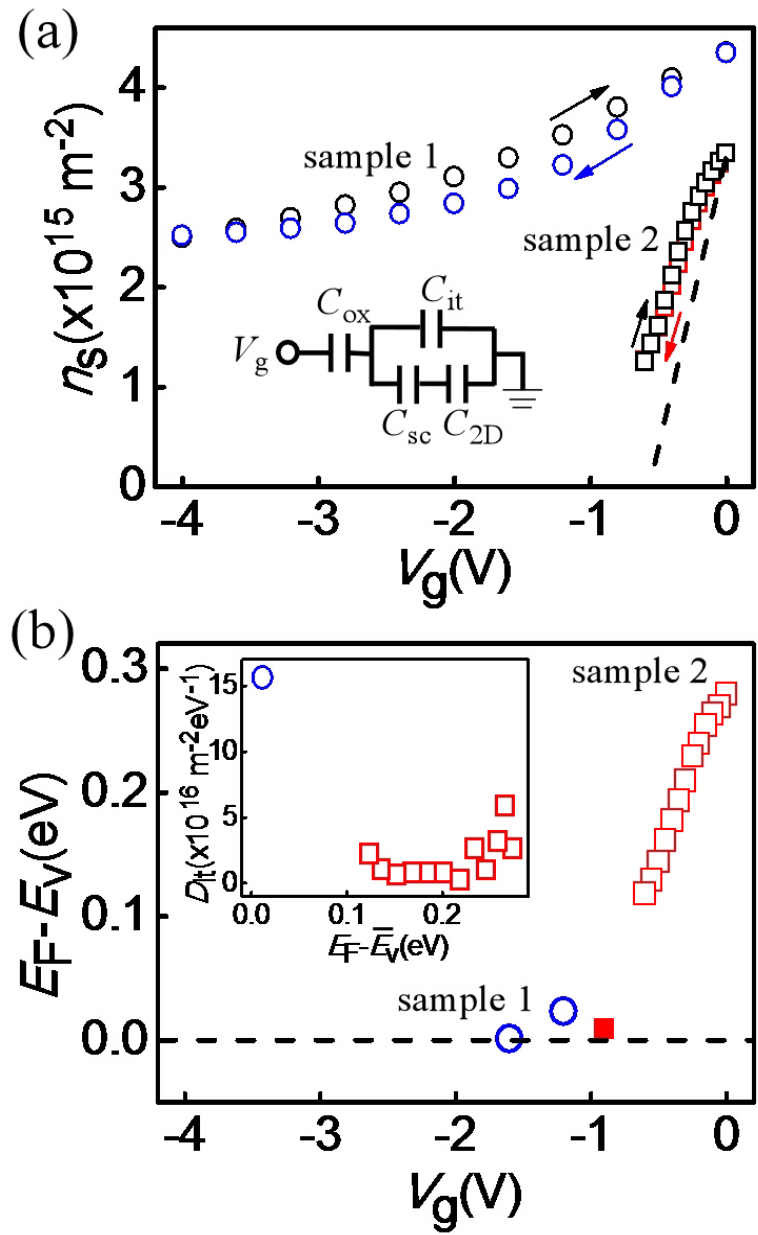

FIG. 2

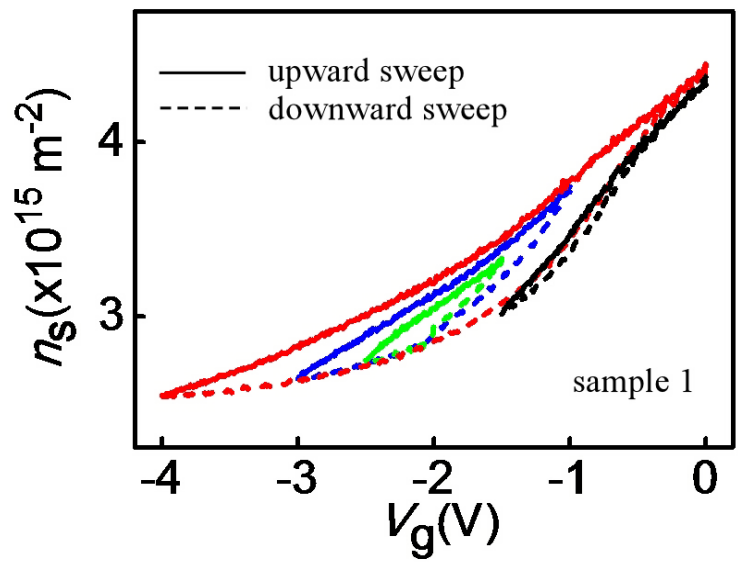

FIG. 3